\documentclass[a4paper,11pt]{article}
\usepackage{epsfig}
\usepackage[utf8]{inputenc}
\usepackage{cite}
\usepackage{hyperref}
\usepackage{epsfig}
\usepackage{amsmath}
\usepackage{amsfonts}
\usepackage{graphicx}
\usepackage{cite}
\usepackage{color}
\usepackage[dvipsnames]{xcolor}
\usepackage{multirow}
\usepackage{amssymb}
\usepackage{graphicx}
\usepackage{caption}

\def\be{\begin{equation}}
\def\ee{\end{equation}}

\begin{document}
\title{{\bf  \LARGE EHT constraint on the ultralight scalar hair \\ of the  M87 supermassive black hole}}
 \author{
{\large Pedro V.P. Cunha}$^{1,2}$,\, 
{\large Carlos A. R. Herdeiro}$^{1,2}$,\, 
{\large Eugen Radu}$^{1}$
\\
\\
$^{1}${\small Departamento de Matem\'atica da Universidade de Aveiro and } \\ {\small  Centre for Research and Development  in Mathematics and Applications (CIDMA),} \\ {\small    Campus de Santiago, 3810-183 Aveiro, Portugal}
 \\
 \\
$^{2}${\small Centro de Astrof\'isica e Gravitaç\~ao - CENTRA, Departamento de F\'isica,}\\ { \small Instituto Superior T\'ecnico - IST, Universidade de Lisboa - UL,}\\ {\small Av. Rovisco Pais 1, 1049-001, Lisboa, Portugal}
}
\date{September 2019}
\maketitle

\begin{abstract}
Hypothetical ultralight bosonic fields will spontaneously form macroscopic bosonic halos around Kerr black holes, via superradiance, transferring part of the mass and angular momentum of the black hole into the halo. Such process, however, is only efficient if resonant: when the Compton wavelength of the field approximately matches the gravitational scale of the black hole. For a complex-valued field, the process can form a stationary, bosonic field-black hole equilibrium state - a black hole with synchronised hair. For sufficiently massive black holes, such as the one at the centre of the M87 supergiant elliptic galaxy, the hairy black hole can be robust against its own superradiant instabilities, within a Hubble time. Studying the shadows of such scalar hairy black holes, we constrain the amount of hair which is compatible with the Event Horizon Telescope (EHT) observations of the M87 supermassive black hole, assuming the hair is a condensate of ultralight scalar particles of mass $\mu\sim 10^{-20}$ eV, as to be dynamically viable. We show the EHT  observations set a weak constraint, in the sense that typical hairy black holes that could develop their hair dynamically, are compatible with the observations, when taking into account the EHT error bars and the black hole mass/distance uncertainty.
\end{abstract}

\newpage

\tableofcontents

\section{Introduction}
The hypothesis that all astrophysical black holes (BHs) when near equilibrium are well described by the Kerr metric~\cite{Kerr:1963ud} - \textit{the Kerr hypothesis} - yields a remarkable scenario. It means that throughout the whole mass spectrum, ranging from solar mass BHs, with $M\sim M_\odot$, all the way until the most supermassive black holes known, with $M\sim 10^{10} M_\odot$, the immense population of astrophysical BHs correspond to the very same object, with only two macroscopic degrees of freedom. One of these is the mass, which merely rescales the BH, leaving a single degree of freedom with impact on the BH phenomenology - the spin. The Kerr hypothesis, therefore, encodes an economical natural order: the landscape of gravitational atoms (BHs) that compose the dark Universe is made up of a single species, varying only in size (by, at least, ten orders of magnitude!) and spin, but otherwise indistinguishable. Such uniformity is a trademark of the microscopic world, where all elementary particles of a given species are indistinguishable, but not of the macroscopic world, where variety is ubiquitous. 

Despite the current lack of tension between observations and the Kerr hypothesis, there are reasons to consider the latter but a fair approximation, within current precision, rather than a fundamental truth. Both fundamental problems  - such as a quantum version of the laws of gravity, and how it impacts on the physics of horizons and classical singularities -, and the phenomenological problem of accounting for the elusive dark matter and dark energy, suggest our current understanding of gravity is incomplete. It may therefore be that the Kerr hypothesis is strictly false for astrophysical BHs at all scales. But an alternative possibility is that the Kerr hypothesis is violated at a higher degree at some narrow interval of scales only, wherein new physics exists, remaining an excellent approximation outside this interval. 

A concrete realisation of the latter possibility is provided by scenarios of hypothetical ultralight bosonic particles, that could constitute part of the dark matter population{~\cite{Suarez:2013iw,Hui:2016ltb,Visinelli:2018utg}}. Inspired by the QCD axion~\cite{Peccei:1977hh}, and with theoretical support in string theory~\cite{Arvanitaki:2009fg}, these scenarios suggest a landscape of such particles, negligibly interacting with standard model constituents, might exist, leaving their gravitational interactions as the only possible smoking gun for their identification. Amongst these, an exciting possibility, is their interaction with BHs, which, could single out a scale (or a range of scales) wherein BHs deviate from the Kerr paradigm. 

Spinning BHs (in particular Kerr BHs) are energy and angular momentum reservoirs that can be classically mined. A very well suited tool for such mining is precisely an ultralight bosonic field. Then, through the phenomenon of superradiance~\cite{Press:1972zz,Brito:2015oca} an appropriate small seed of such field (provided, say, as a quantum fluctuation) will grow into a macroscopic condensate of bosonic particles - a Bose-Einstein condensate - storing a non-negligible fraction of the original BH mass and angular momentum. When the energy/angular momentum transfer from the BH to the bosonic halo saturates, the corresponding BH-halo  state may or may not be stationary. If the bosonic field is real, the BH-halo system emits gravitational radiation and slowly decays back to a Kerr BH~\cite{Arvanitaki:2010sy}. But if the bosonic field is complex, the BH-halo system is stationary, in fact a hairy BH - dubbed BHs with synchronised hair~\cite{Herdeiro:2014goa}. 

BHs with synchronised hair are not absolutely stable. They are themselves prone to their own superradiant instabilities~\cite{Herdeiro:2014jaa,Ganchev:2017uuo,Degollado:2018ypf}. However, the timescale of these instabilities is larger than the one of the initial Kerr superradiant instability that formed the hair, and, in the right mass range it becomes cosmologically large. A suggestive possibility is then the following. A Kerr BH develops ultralight bosonic hair in an astrophysical time scale; the hairy BH is then effectively stable, since it is superradiantly unstable only in a cosmological timescale. This turns out to be a realisable scenario for supermassive BHs, such as the one recently observed by the Event Horizon Telescope (EHT) collaboration~\cite{Akiyama:2019cqa,Akiyama:2019fyp,Akiyama:2019eap} at the centre of the supergiant elliptic galaxy M87, henceforth referred to as the M87 BH.

The scenario in this paper is therefore that the M87 BH is hairy, due to an appropriate ultralight scalar field. Appropriate means its mass is in the right range to make the superradiant instability of the original Kerr BH grow in a sufficiently small fraction of the Hubble time, yielding a hairy BH that is stable in a cosmological time scale. Since the shadows of a Kerr and a hairy BH with the same total mass and angular momentum differ~\cite{Cunha:2015yba}, and since the EHT observation is compatible with the M87 BH being of Kerr type, we shall then inquire how much the EHT observations constrain the hair. As we shall see, for the most interesting mass ranges, as to make the hairy BH dynamically viable, the EHT constraint is weak, and it is essentially compatible with a hairy BH that could have dynamically formed from superradiance and it is in a long lived, albeit not absolutely stable, state. 

This paper is organised as follows. In section~\ref{section2} we discuss the physical scenario under which a BH with synchronised scalar hair could form from superradiance and be effectively stable within a cosmological timescale. In section~\ref{section3} we describe the part of the domain of existence of hairy BHs that is dynamically viable, according to the criteria in section~\ref{section2} and that we shall study in the remaining of the paper. In section~\ref{section4} we consider the Kerr BH shadow and we obtain an approximate expression for the shadow areal radius, valid for arbitrary observation angle and dimensionless spin value. In section~\ref{section5} we analyse the shadows of the hairy BHs in the relevant domain and obtain an approximate expression for the areal shadow radius, in terms of that of a comparable Kerr BH, $i.e.$ with the same mass,  and a parameter measuring the hairiness of the BH. In section~\ref{section6} the expression obtained in section~\ref{section5} is applied to the case of the M87 supermassive BH. Then, considering the EHT observational errors, together with the errors in the mass estimate, we constrain the hairiness compatible with the observations. Brief final remarks are given in Section~\ref{section7}. {Unless otherwise stated, natural units $c=1=G=h$ are used.}

\section{The hair formation and hair instability timescales}
\label{section2}
The timescale of BH superradiance is extremely sensitive to the occurrence, or not, of a resonance between the gravitational scale of the BH and the Compton wavelength of the ultralight particles. Consider a massive, complex, scalar field, $\Phi$, described by the Klein-Gordon equation, $\Box \Phi=\mu^2 \Phi$, with mass $\mu$, on the background of a Kerr BH with mass $M$. Maximal efficiency occurs for $M\mu\simeq 0.4$~\cite{Dolan:2007mj}  and for a spin close to extremality; for the M87 BH, for which we take $M_{\rm M87}\sim 6\times 10^9 M_\odot$\footnote{For the considerations in this section this approximate value suffices. More accurate values will be considered in section~\ref{section6}. This value is suggested from stellar dynamics~\cite{Gebhardt:2011yw} and favoured by the EHT observations~\cite{Akiyama:2019cqa}. A value half of this is suggested by gas dynamics~\cite{Walsh:2013uua}. The spin of the M87 BH is largely unknown, with different claims in the literature, see $e.g.$~\cite{Nokhrina:2019sxv,Tamburini:2019vrf}.} this resonant scalar field mass is
\begin{equation}
\mu_{\rm r}=\frac{0.4}{M_{\rm M87}} \simeq 4\times 10^{-20} \  {\rm eV} \ .
\label{eq1}
\end{equation}
At maximal efficiency, the e-folding time of the superradiant instability's leading mode is~\cite{Dolan:2007mj} 
\begin{equation}
\Delta t\sim 10^7 \tau_{LC} \ ,
\end{equation}
where $\tau_{LC}$ is the light crossing time of the BH. For the M87 BH,  $\Delta t\sim 10^4$ years. We call this the \textit{hair formation timescale}. This means that if an ultralight boson of mass $\mu_{\rm r}$ exists, the M87 BH would develop scalar hair in an astrophysically short time scale. 

If the resonance $\mu=\mu_{\rm r}$ is missed, however, this timescale grows extremely fast: as $(M\mu)^{-8}$~\cite{Detweiler:1980uk} for the leading mode and $M\mu\ll 1$; or as $10^{7}e^{3.7M\mu}$ for $M\mu\gg 1$~\cite{Zouros:1979iw,Arvanitaki:2010sy}. In other words, if $M\mu$ misses the resonant sweet spot by one order of magnitude, either above or below, the timescale of the leading mode of the superradiant instability of Kerr becomes considerably larger than the Hubble time, and the Kerr BH does not become hairy.  On the lower end, the hair formation timescale becomes larger than one tenth of the Hubble time for $M\mu<0.05$~\cite{Degollado:2018ypf}, for the M87 BH mass. Thus, a conservative bound on the formation timescale is to take $M\mu>0.1$, when the formation timescale becomes around a thousandth of the Hubble time.

Supermassive BHs in matter rich environments, such as galactic centres, are expected to grow in time. Thus, a supermassive BH  with $\sim 10^9 M_\odot$, such as the one at the centre of M87, will have evolved, due to accretion and mergers, from one (or many) BHs with mass several orders of magnitude lower - see $e.g.$~\cite{Volonteri:2002vz}. Only when the BH grows to the size that resonates with $\sim \mu_{\rm r}$ does the superradiant energy/angular momentum extraction becomes efficient producing a sufficiently non-Kerr BH. At all other scales BHs remain Kerr-like. 

Once the hairy BH forms, one must consider its leading superradiant instability. The leading instability has an e-folding time - dubbed \textit{hair instability timescale} - larger than the Hubble time if $M\mu\lesssim 0.25$~\cite{Degollado:2018ypf}. Thus, a Kerr BH with mass $M=M_{\rm M87}$ becomes hairy in an astrophysical timescale and the hair is stable in a cosmological timescale if {(conservatively rounding off $0.25$ to $0.3$)}
\begin{equation}
\label{range}
\mu M_{\rm M87}\in [0.1,0.3] \ \ \ \Rightarrow  \ \ \ \mu\in \textrm{{$[1\,,\,3]$}} \times 10^{-20} \  {\rm eV} \ .
\end{equation}

{We remark that there is a dependence on the BH dimensionless spin parameter $a$ in determining the resonant scalar field mass $\mu_{\rm r}$, although the instability is considerably more sensitive to $M\mu$ than $a$. If the spin is not near extremal, this changes the ideal value of $M\mu$ given in (\ref{eq1}), which could push down slightly, but not significantly, the lower end value of the interesting mass range given in (\ref{range}) (see also discussion in Section~\ref{section7}).
}

How much energy can be extracted from the Kerr BH region into the hair? Fully non-linear numerical evolutions of the superradiant instability of a Kerr BH triggered by a complex \textit{vector} field were performed in~\cite{East:2017ovw}, leading to the formation of that BH with synchronised (vector)  hair~\cite{Herdeiro:2017phl}, first discussed in~\cite{Herdeiro:2016tmi}. These simulations showed that the maximal energy extracted was $\sim 0.09M$ of the original BH with mass $M$. The largest energy extraction, moreover, occurred for the lowest values of $M\mu$ ($M\mu=0.25$, for the simulations in~\cite{East:2017ovw}), for which the superradiant growth was slower. The trend moreover suggests that the $9\%$ may be close to the maximal possible value, in the vector case. The corresponding value in the scalar case is unknown. Since the process is slower in the scalar case, it is conceivable that larger energy extractions are possible - see~\cite{Brito:2014wla}. But, in any case, thermodynamic sets a limit of 29\% to the rotational energy that can be extracted from a Kerr BH.

\section{The selected part of the domain of existence}
\label{section3}
The full domain of existence of BHs with synchronised scalar hair was obtained in~\cite{Herdeiro:2014goa,Herdeiro:2015gia}.  These are stationary and axisymmetric solutions of the Einstein-(massive, complex)Klein-Gordon system. The shadows of these BHs have been explored in~\cite{Cunha:2015yba,Cunha:2016bjh,Vincent:2016sjq}. However, a detailed study of the shadow properties in the region of dynamical interest unveiled in the previous section remains to be done.  The goal of the reminder of this paper is to perform such study and relating it to the EHT observations. 

The part of the domain of existence describing the astrophysically viable solutions, in relation to the M87 BH, as described in the last section, corresponds to values of $M\mu$ in accordance to~\eqref{range}. For $M\mu\lesssim 0.1$, obtaining numerically the hairy BH solutions becomes challenging, due to the different scales involved in the problem. So, we shall perform our analysis of the shadows in a section of the domain of existence for $0.2\leqslant M\mu \leqslant 0.5${, which allowed us to obtain interpolations with more data. The main conclusions are not substantially affected by this choice of sample.} As we shall see, the obtained trend is already informative.

\begin{figure}[h!]
\begin{center}
\includegraphics[width=0.55\textwidth]{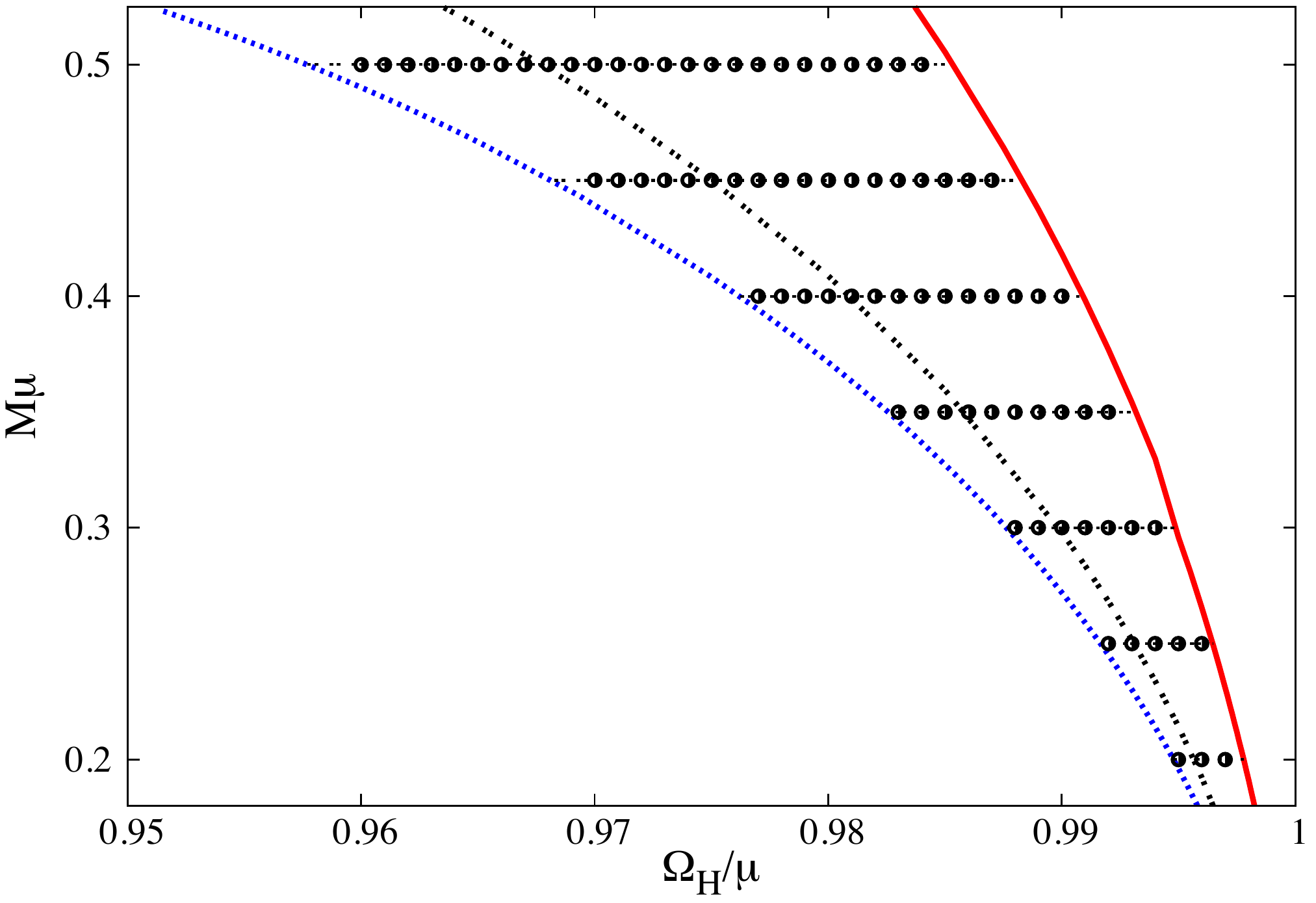}
\caption{\small The section of the domain of existence of hairy BHs to be analysed. We have chosen sequences of representative solutions with constant $M\mu$ - black dots. Their shadow properties are analysed and the corresponding trends interpolated for the whole region. The dashed dotted line separates solutions with more (to the right) and less (to the left) than 29\% of the spacetime energy in the scalar hair.}
\label{HBHs}
\end{center}
\end{figure}

In Fig.~\ref{HBHs} we exhibit the configurations analysed to obtain the behaviour of the shadows of the hairy BHs in the dynamically viable region.  
They are displayed in the configuration space $(\Omega_H/\mu,\,M\mu)$,  with $\Omega_H$ and $\mu$ being, respectively, the BH horizon angular velocity and the boson mass. The synchronisation condition\footnote{{The synchronization condition is an equilibrium requirement on the existence of these hairy BH solutions ($i.e.$ within the valid domain in Fig.~\ref{HBHs}), and it is not directly used in the rest of the analysis.}} means that $\omega=\Omega_H$ for this family of solutions, where $\omega$ is the frequency of the complex scalar field, that oscillates harmonically (but with a time independent energy momentum tensor). The Kerr limit, in which the hair vanishes, is provided by the dotted blue line; the rotating boson star limit, in which the horizon vanishes, is provided by the solid red line.

As an alternative to the $\{\Omega_H/\mu,\,M\mu\}$ labelling of this section of the domain of existence of the hairy BHs, each solution can also be labeled by the pair $\{p,M\!\mu\}$, where~\cite{Delgado:2016zxv}, 
\begin{equation}
p\equiv 1-\frac{M_H}{M} \ ,
\end{equation}
$M_H$ is the horizon energy (measured by a Komar integral) and $M$ the ADM energy. Thus, $p$ measures the fraction of the spacetime energy in the hair. This parameter satisfies $0\leqslant p \leqslant 1$; $p=0$ ($p=1$) corresponds to the Kerr (boson star) limit, displayed as the dotted blue (solid red) lines in Fig.~\ref{HBHs}. In the figure, a dotted-dashed black line denotes the $p=0.29$ threshold, above which the hairy BHs cannot form from the superradiant instability of Kerr. 
{Nevertheless, configurations with $p>0.29$ are also equally valid equilibrium solutions to the Einstein-Klein-Gordon field equations. Although at the moment this is unclear, these configurations could perhaps form by alternative channels other than the superradiant instability. For the analysis being performed, considering configurations with $p>0.29$ increases the sample size for the shadow interpolations in Section~\ref{section5}.}

\section{The Kerr BH shadow}
\label{section4}
{The calculation of the BH shadow for a Schwarzschild and Kerr BH was introduced by Synge~\cite{Synge:1966okc} and Bardeen~\cite{Bardeen1973} respectively}. A pioneering computation in an astrophysical environment was done by Luminet~\cite{Luminet:1979nyg,Luminet:2019hfx} and its measurability was first assessed in~\cite{Falcke:1999pj} - see~\cite{Cunha:2018acu} for a review and~\cite{Cunha:2016bpi} for a working setup to compute shadows via  ray tracing. 

Given an observational setup, the observer can measure the BH shadow image\footnote{{The shadow in the image domain is rescaled with respect to its angular size $\vartheta$ by a factor $\mathcal{R}$, see \eqref{angularsize}.}} area $\mathcal{A}$. We define  the shadow \textit{areal radius} as 
\begin{equation}
S\equiv \sqrt{\frac{\mathcal{A}}{\pi}} \ ,
\end{equation}
 which is well defined even for non-circular shadow shapes. In what follows, the shadow radius\footnote{{Other possible measures for the shadow size can be found in the literature, $e.g.$ see~\cite{Grenzebach:2015oea}}} $S$ will be compared between hairy and Kerr BHs. The latter is known analytically in particular cases, as we shall now review for our subsequent application.

\subsection{Two cases for which the Kerr shadow areal radius is exactly computable}
For an observer at infinity, the Kerr shadow edge is known analytically for all spin values $a=J/M$~\cite{Cunha:2018acu}, where $J$ is the total angular momentum of the spacetime and the existence of a horizon requires $0\leqslant a^2\leqslant M^2$. From this analytic knowledge of the shadow edge, however, it might not always be possible to compute $\mathcal{A}$ exactly, and hence the Kerr shadow areal radius, denoted $S_{\rm Kerr}(a,\theta_o)$, which generically depends on $a$ and the polar angle of the observer $\theta_o$. But in two special cases this is possible. 

The first case is when the observer is on the rotation axis, $i.e.$ when $\theta_o=\{0,\pi\}$. Then, the Kerr shadow edge is circular due to axial-symmetry. In this case the shadow radius $S_{\rm Kerr}(a,{\rm axis})$ is fully determined by a zero angular momentum spherical photon orbit with a Boyer-Linquist radial coordinate $r$. For this special case $\mathcal{A}$ can be obtained exactly and so can the shadow areal radius. Using the results in~\cite{Teo2003}, the latter is obtained to be, as a function of $a$:
\begin{equation}
S_{\rm Kerr}(a,{\rm axis})= \sqrt{\chi +a^2} \ ,
\end{equation}
where
\[ \chi=r^2\,\left(\frac{3r^2 +a^2}{r^2-a^2}\right) \ ,\ \ \ \ \  \frac{r}{M}=1+2\sqrt{1-\frac{a^2}{3}}\cos\left(\frac{1}{3}\arccos\left[\frac{1-a^2}{\sqrt{\left(1-\frac{a^2}{3}\right)^3}}\right]\right) \ .
\]
Observe that for $a=0$, then $r=3M$, $\chi=27M^2$ and $S_{\rm Kerr}(0,{\rm axis})=\sqrt{27}M$, which are the familiar Schwarzschild light ring coordinate, the corresponding impact parameter (squared) and shadow areal radius.

The second case is for an extremal Kerr BH ($i.e.$ $|a|=M$), viewed from the equatorial plane, $i.e.$ with $\theta_o=\pi/2$. Then, the shadow edge is not circular. However, the shadow shape $y(x)$ in Cartesian-like coordinates $(x,y)$ simplifies into \cite{Cunha:2018acu} (in units of $M$):
\[y(x)=\pm\sqrt{11+2x-x^2+8\sqrt{2+x}} \ ,\qquad x\in[-2\,,\,7] \ ,\]
in which case, the area $\mathcal{A}$ can be explicitly computed:
\[\mathcal{A}=\int_{-2}^7 2\,y(x)\,dx=(15\sqrt{3}+16\pi)M^2 \ ,\]
which leads to a shadow radius 
\begin{equation}
S_{\rm Kerr}\left(\pm M,\frac{\pi}{2}\right)=\sqrt{16+\frac{15\sqrt{3}}{\pi}} M\ .
\end{equation}

\subsection{An approximation for the areal radius of the Kerr shadow}

We were not able to find an exact expression for the Kerr shadow areal radius in the generic case.   We have verified, however, that as seen by an observer at infinity, $S_{\rm Kerr}(a,\theta_o)$ can be estimated (within an error $\lesssim 0.8 \%$) as:\footnote{{This error was determined through a direct comparison of the approximate formula with the corresponding Kerr values. Recall that the Kerr shadow edge is known analytically and determining the shadow areal radius amounts to solving, numerically the area integral. Thus the Kerr shadow areal radius, albeit obtained numerically, is computed with a precision considerably better than 0.8\%.}}
\begin{equation}
S_{\rm Kerr}(a,\theta_o)\simeq S_{\rm Kerr}(a,{\rm axis}) + \frac{2|a|\theta_o}{\pi M}\left[S_{\rm Kerr}\left(M,\frac{\pi}{2}\right)-S_{\rm Kerr}(M,{\rm axis})\right] \ ,
\end{equation}
where $S_{\rm Kerr}(M,{\rm axis})=(2+2\sqrt{2})M$. This approximation will be used in the following.

To make contact with the Kerr limit in the domain of existence displayed in Fig.~\ref{HBHs}, we observe that from the points along the blue dotted line therein,  the Kerr spin $|a|$ can be obtained from $M\!\mu$ using
\[a=\frac{M^2\Omega_H}{M^2\Omega_H^2 +1/4}\,\,,\qquad M\Omega_H\simeq b_1 + b_2\,M\mu + b_3\,M^2\mu^2\ ,\]
where the first expression is exact for Kerr BHs, and the second is a good approximate relation along the Kerr existence (blue) line in Fig.~\ref{HBHs}, with parameters
\begin{equation}
(b_1,b_2 ,b_3)=(-0.00926172,1.08238,-0.209874) \ .
\end{equation}
Thus, for the Kerr BHs in Fig.~\ref{HBHs}, the shadow areal radius becomes a function of $M\mu$ and $\theta_o$, $S_{\rm Kerr}(a(M\mu),\theta_o)$.

\section{Hairy BHs shadow in the considered domain of existence}
\label{section5}

Since the shadows of the hairy BHs were obtained through a numerical ray tracing procedure~\cite{Cunha:2016bpi}, they correspond to an observer at finite, rather than infinite, distance from the BH. The observer is placed at a finite perimetral distance $\mathcal{R}=\sqrt{g_{\varphi\varphi}}(r_o,\pi/2)$, where $\partial/\partial\varphi$ is the Killing vector field associated to axial-symmetry. The quantity $\mathcal{R}$ is defined for each radial coordinate $r_o$. For two observers $\{\mathcal{O}_1,\mathcal{O}_2\}$, respectively with $\{\mathcal{R}_1,\mathcal{R}_2\}\gg M$ and corresponding shadow radii $\{S_1,S_2\}$, a simple extrapolation can provide a good approximation for the shadow radius $S_\infty$ at infinity:
\[S_\infty\simeq S_2 - \left(\frac{S_2-S_1}{1-\mathcal{R}_2/\mathcal{R}_1}\right).\]
In our setup, we take $\mathcal{R}_1=100M$ and $\mathcal{R}_2=200M$.

\subsection{Hairy BH shadow approximation for $\theta_o=17º$}
\begin{figure}[h!]
\begin{center}
\includegraphics[width=0.55\textwidth]{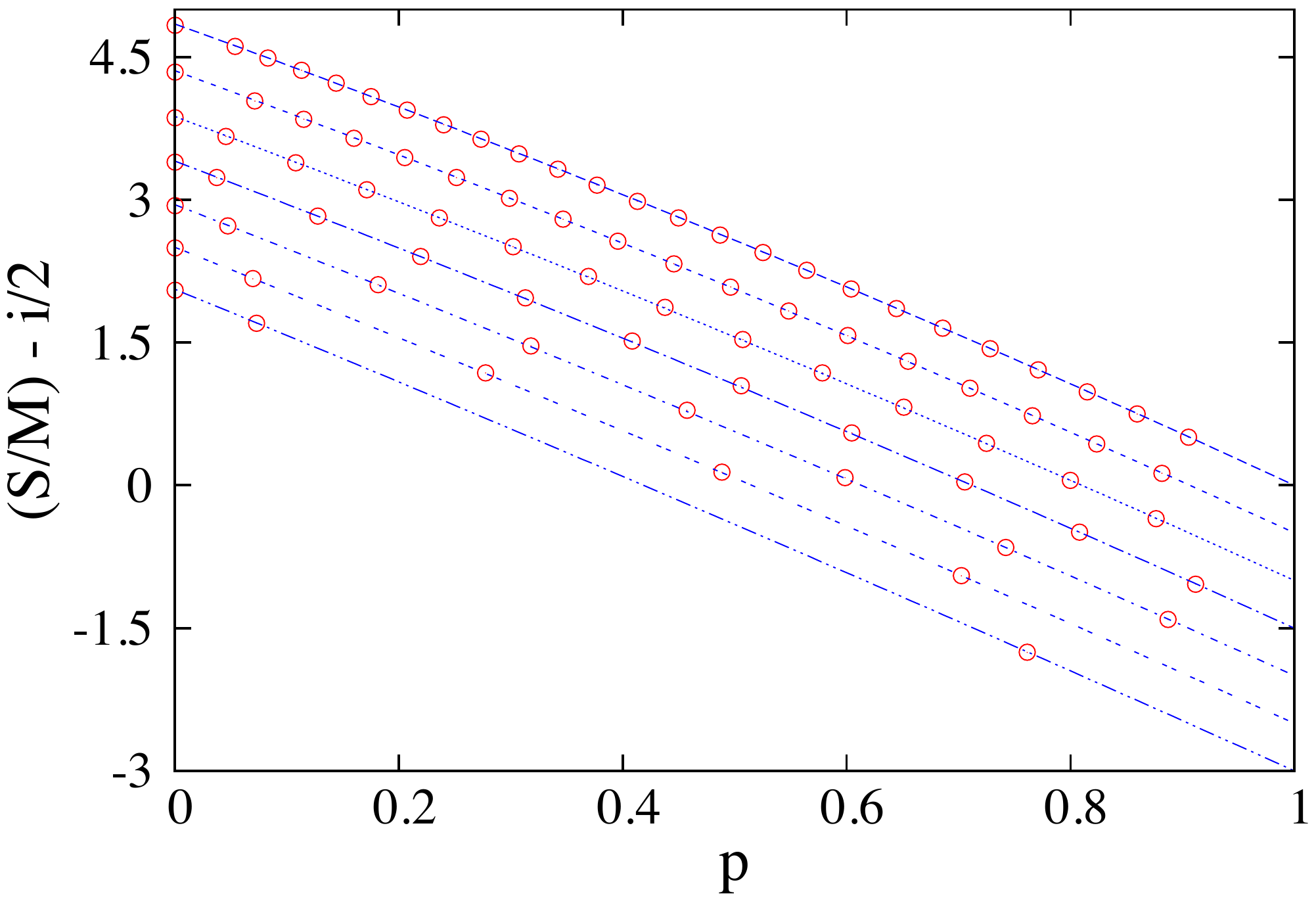}
\caption{\small Areal shadow radius for hairy BHs and the analytic approximation~\eqref{hsa}. Circles correspond to data for the individual solutions in Fig.~\ref{HBHs}.  Each straight line is a set with constant $M\mu$  in Fig.~\ref{HBHs}, given by $M\mu=0.5-i/20$ and $i=\{0,\cdots,6\}$ as we go from the top to the bottom straight line. The function exhibited in the $y$-axis is $S_{\rm hairy}(p,M\!\mu,17º)/M-i/2$, as to more clearly distinguish the different lines, where $i$ is a function of $M\mu$ via $i=10\times(1-2M\mu)$. }
\label{shadow-data}
\end{center}
\end{figure}

The angle between the M87 BH spin and the line of sight has been estimated to be $17º$ from the observed jet~\cite{2018ApJ...855..128W}. Choosing, then,  $\theta_o=17º$ in order to compare with EHT's M87 observation, the shadow's areal radius at infinity of the hairy BHs, $S_{\rm hairy}(p,M\!\mu,\theta_o)$  can be approximated within an error $\lesssim 0.8 \%$ as:
\begin{equation}
S_{\rm hairy}(p,M\!\mu,17º)\simeq (1-p)\Bigg(S_{\rm Kerr}(a(M\!\mu),17º)\,+\,\beta_1p\,\,M^2\!\mu \Bigg)\ ,\qquad\quad \beta_1\simeq 1.21455 \ ,
\label{hsa}
\end{equation}
The accuracy of this approximation is clear from~Fig.~\ref{shadow-data}, where both the shadow radius of the hairy BHs and the function $S_{\rm hairy}(p,M\!\mu,17º)$, are exhibited.

\begin{figure}[h!]
\begin{center}
\includegraphics[width=0.55\textwidth]{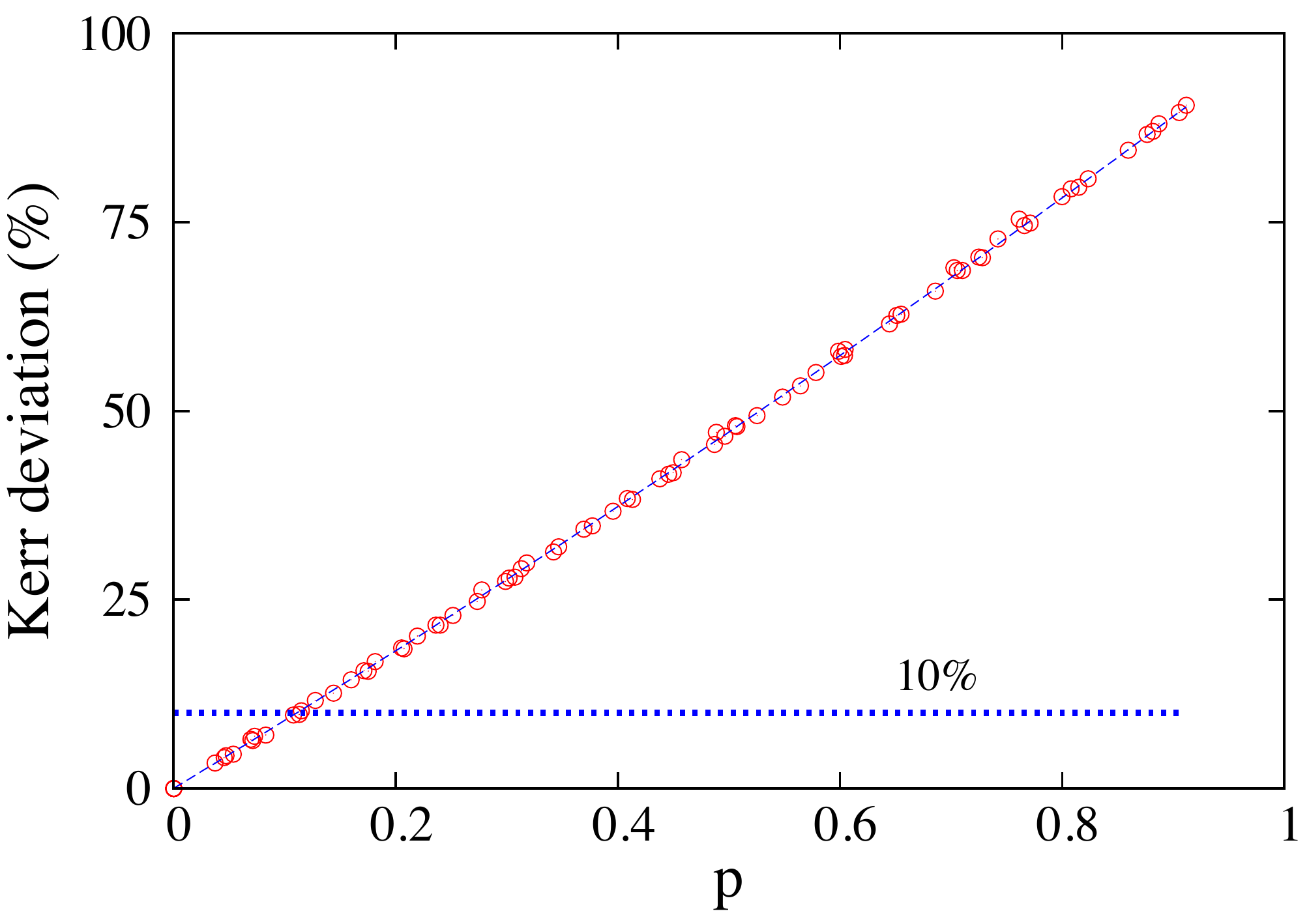}
\caption{\small Deviation $\delta S$ between the hairy BHs shadow areal radius and that of Kerr BHs, as function of $p$, for $\theta_o=17º$. The fit function $p + A\,p(p-1)$ with $A\simeq 0.111159$ captures the main features of $\delta S$.}
\label{deviation}
\end{center}
\end{figure}

The analysis of the hairy BHs data, leading to~\eqref{hsa}, shows that the shadow's areal radius relative deviation $\delta S$, between hairy and Kerr BHs, depends very weakly on $M\!\mu$. It is therefore accurately  parameterized by a function of $p$ only:
\begin{equation}
\hspace{1cm}
\delta S(p)\equiv 1-\frac{S_{\rm hairy}(p,M\!\mu,17º)}{S_{\rm Kerr}(a(M\!\mu),17º)}\simeq p + p(p-1)A\ ,\qquad\qquad\textrm{with}\quad A\simeq 0.111159\ .
\label{eq-psi}
\end{equation}
This approximation is represented in Fig.~\ref{deviation} as a solid line, together with the corresponding data (red circles), showing a very good agreement. 

A rough conclusion from this analysis is that a Kerr deviation no larger than  $\sim$10\%  is compatible with a $p\lesssim$ 11\% (dotted line in Fig.~\ref{deviation}). Indeed, the EHT measurement of the M87 BH shadow has an error bar of around 10\% as discussed in the next section where a more precise statistical analysis is performed. To make contact with the observations, we note that 
for a dimensionless areal radius $S/M$, the corresponding angular radius in the sky is 
\begin{equation}
\vartheta=(S/M)\frac{M}{\mathcal{R}} \ .
\label{angularsize}
\end{equation}
This relates theory with observation. Using it we will  now restrict the values of $p$ of a hairy BH that may be compatible with the EHT M87 BH observation data.

\section{Application to the M87 BH shadow}
\label{section6}

The EHT observation measures the emission ring \textit{diameter} to be 42$\pm 3 \ \mu$as\cite{Akiyama:2019cqa}. As discussed in~\cite{Akiyama:2019eap}, this emission ring diameter is not simply assumed to be the edge of the shadow. Calibration between the emission ring diameter and the photon ring (determining the edge of the shadow) based on GRMHD simulations leads to a 10\% offset between the two. 
{Although this offset is estimated from Kerr, it is conceivable that a similar effect arises for hairy BHs in the region of interest of Fig.~\ref{HBHs}, which are still very much Kerr-like.}
 Thus, we assume the M87 BH shadow diameter is 10\% smaller than the EHT's observed emission ring, leading to an observed angular size of the shadow (corresponding to the areal \textit{radius}) of
\begin{equation}
\vartheta_o=\textrm{{$\left(\,18.9\pm 1.5 \right)$}}\ \mu{\rm as} \  . 
\label{oa}
\end{equation} 

\begin{figure}[ht!]
\begin{center}
\includegraphics[height=6.2cm]{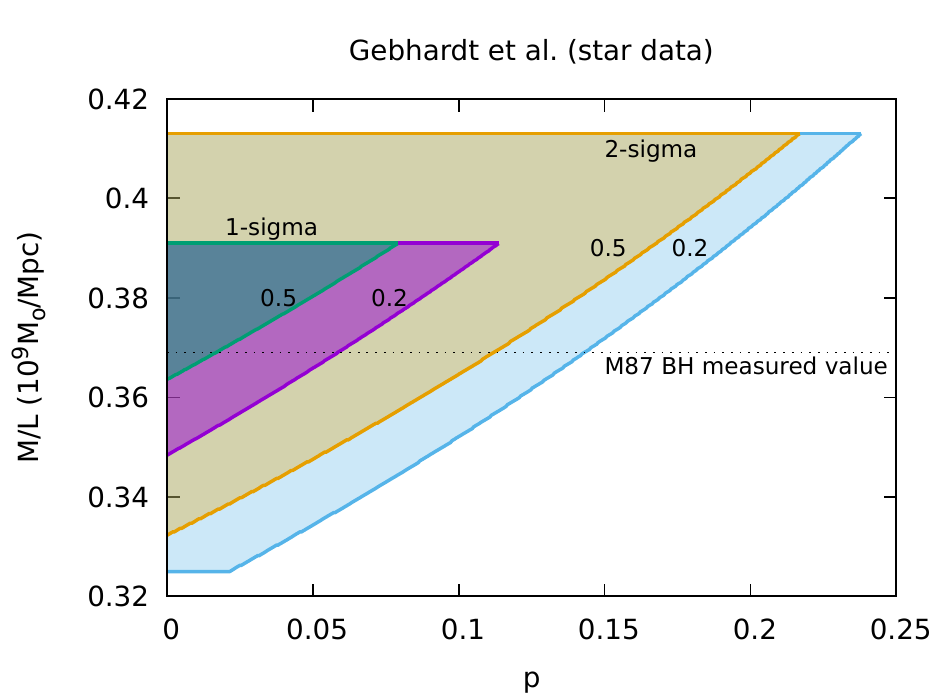}\includegraphics[height=6.2cm]{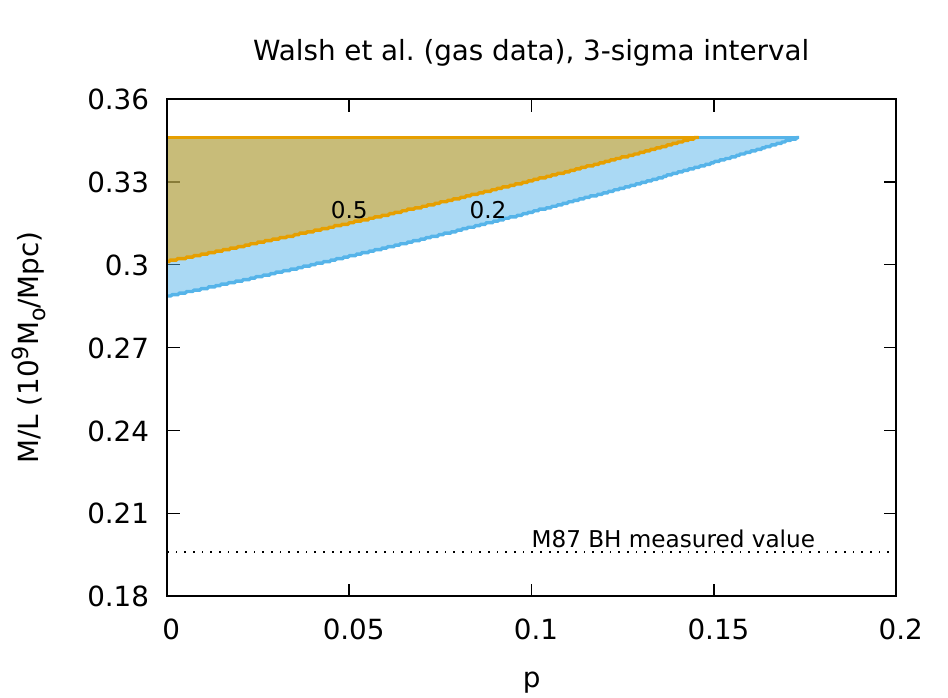}
\caption{\small $(p,\lambda=M/L)$ domain providing values of $\vartheta =\lambda S_{\rm hairy}(p,M\!\mu,17º)/M$  consistent with the EHT observation. (Left panel) within 1 and 2 standard deviations for the star motion data. (Right panel): within 3 standard deviations for the gas data. Small numbers indicate value of $M\!\mu$.}
\label{img-sigma}
\end{center}
\end{figure}

The error bars in~\eqref{oa} already provide some margin to accommodate a non-Kerr BH with the same mass. But a proper analysis must, in addition, take into account the error in the mass measurement, which must be an independent measurement from the EHT observations. As discussed in~\cite{Akiyama:2019eap}, both the independent measurements of the M87 BH mass, by Gebhardt
et al.~\cite{Gebhardt:2011yw} based on star dynamics, and Walsh et al.~\cite{Walsh:2013uua} based on gas motion, actually directly measure the ratio $\lambda= M/L$, rather than $M$, where $L$ is the luminosity distance, that we identify with $\mathcal{R}$. The mass in these works is then obtained \textit{assuming} $L$ = 17.9 Mpc, since the relative error for the distance is smaller. Taking their reported values for M and inferring the associated ratios for that distance, one has:
\[Gebhardt \ et \ al. \ \textrm{(star motion):}\qquad M=(6.6\pm 0.4)\times 10^9 M_\odot\ ,\qquad \lambda= 0.369\pm 0.022 \left(\frac{10^9 M_\odot}{\textrm{Mpc}}\right) \ ,\]
\[
Walsh \ et \ al. \  \textrm{(gas motion):}\qquad M=3.5^{+0.9}_{-0.7} \times 10^9 M_\odot\ ,\qquad \lambda= 0.196^{+0.05}_{-0.04} \left(\frac{10^9 M_\odot}{\textrm{Mpc}}\right)\ .
\]
Choosing either of these data sets, we can now analyse the domain in the $(p,\lambda)$ plane that provides an angular shadow size; using~\eqref{hsa} and~\eqref{angularsize} yields 
\begin{equation}
\vartheta = \lambda \frac{S_{\rm hairy}(p,M\!\mu,17º)}{M}\simeq \lambda  (1-p)\Bigg(\frac{S_{\rm Kerr}(a(M\!\mu),17º)}{M}\,+\,\beta_1p\,\,M\!\mu \Bigg)\ ,
\end{equation}
consistent with the EHT shadow, within a certain number of standard deviations. This analysis is performed in  Fig.~\ref{img-sigma}. 
The left panel considers the star motion data. The shaded regions encode values of  $(p,\lambda)$ consistent with the EHT shadow observation, within 1 and 2 standard deviations, for the limiting values of $M\mu$ in the sample of solutions analysed. We conclude that within 1 standard deviation, $0<p<0.12$, and within 2 standard deviations $0<p<0.24$, for $M\!\mu=0.2$. Slightly more restrictive values hold for $M\!\mu=0.5$. The right panel considers the gas data. In this case, values in the $(p,\lambda)$ domain exhibited can only agree with the EHT observation within 3 standard deviations of the observation error.

Considering the star dynamics data within one standard deviation, a hairy BH with $p\lesssim 0.12$ is compatible with the EHT observations. The trend with $M\mu$ in Fig.~\ref{img-sigma} indicates, moreover, that for even lower values of $M\mu$ -- recall~\eqref{range} -- the accommodated $p$ is even slightly larger. Taking the simulations with a vector field discussed in Section~\ref{section2} as an estimate of how much energy could be extracted dynamically into the hair, the tentative conclusion (with the caveat that the precise maximum amount of energy extractable dynamically is unknown in the scalar case) is that dynamically viable hairy BHs are compatible with the EHT observations, given the error bars. 

The gas data, on the other hand, disfavours the hairy BHs, which is manifest in the mostly empty right panel of Fig.~\ref{img-sigma}, but it is also at some tension with the Kerr model, from the EHT observations. Within two standard deviations the data is incompatible with the model.

\section{Final remarks}
\label{section7}

With the advent of the first observation of a BH shadow by the EHT collaboration~\cite{Akiyama:2019cqa,Akiyama:2019fyp,Akiyama:2019eap}, a new direct window has now been opened into the strong gravity regime surrounding BHs. Together with the recent breakthroughs in gravitational wave astrophysics~\cite{Abbott:2016blz,LIGOScientific:2018mvr}, and the precision upgrades that are expected to follow, the shadow observation opens the tantalizing possibility of testing existing BH models with direct observations. 

In this paper, we have considered the possibility the M87 supermassive BH has ultralight synchronised hair. We have made the case that some of these hairy BHs could be dynamically viable as a model of such a supermassive BH. Moreover, we have shown that the current EHT data, when taken together with the most favoured independent measurement of the mass of the M87 BH is compatible with the estimated range for dynamically viable hair. Thus such a hairy BH could be mistaken by a Kerr BH within all current measurements. See~\cite{Davoudiasl:2019nlo,Hui:2019aqm,Chen:2019fsq,Bar:2019pnz,2018ApJ...855..128W,Roy:2019esk} for other constraints on ultralight dark matter from EHT data and, $e.g.$~\cite{Bambi:2019tjh,Tian:2019yhn,Vagnozzi:2019apd,Contreras:2019cmf} for the impact of these data on other scenarios for M87. 

{Our analysis contains assumptions and some possible caveats, including:
\begin{description}
\item[$\bullet$] In eq.~\eqref{eq1} we considered the resonant mass corresponding to the most efficient superradiant scenario, which in particular assumes a near extremal Kerr BH.  If the spin is not near extremal ($i.e.$ ideal to make superradiance as efficient as possible) this changes the ideal value of $M\mu$ given in eq. (1) in the text and, most importantly, it reduces the efficiency of the process and increases the timescale - see Fig. 6 of~\cite{Dolan:2007mj}.  As the dimensionless spin of the Kerr BH varies from $0.5$ to $0.999$, the timescale at maximal efficiency can vary by almost four orders of magnitude. This still allows the formation of scalar hair in less than 1\% of a Hubble time in the M87 case: for maximal efficiency the time scale was $10^4$ years for the M87 mass.  This variation in the most efficient $M\mu$ could push down slightly, but not significantly, the lower end value of the interesting mass range given in eq.~\eqref{range}.
\item[$\bullet$] Although we have motivated that the most interesting mass interval in the context of our analysis is given by~\eqref{range}, the analysis of hairy BH solutions was performed in a different mass range, $cf.$ Fig.~\ref{HBHs}. This was justified in Section~\ref{section3} and we believe the main conclusions are not substantially affected by this choice of sample.
\item[$\bullet$] Our work assumes the scalar hair around M87 is truly stationary, described by a minimally coupled massive, complex scalar field and forms from superradiance. If other mechanisms can form hairier BHs, or  for other sorts of BHs with scalar hair (even if only approximately stationary), our conclusions do not apply, as, for example, in the scenario discussed in~\cite{Clough:2019jpm,Hui:2019aqm}. 
\item[$\bullet$]  In this paper we have used a single number (the shadow aerial radius) to set constraints. Other shadow measures could also be introduced ($e.g.$ shadow deviation from a circle). However, due to the precision of the EHT measurement, such quantities would be too poorly constrained, at the moment. Such an analysis will be certainly interesting when more precise observations become possible.
\item[$\bullet$] We have assumed that the M87 BH spin makes an angle of  $17º$ with the line of sight, as suggested from the jet~\cite{2018ApJ...855..128W} and also assumed by the EHT analysis. 
\item[$\bullet$] We have assumed that there is an offset of about 10\% between the size of the photon ring and the emission ring observed by the EHT. For Kerr this is justified by numerical GRMHD simulations - see also~\cite{Narayan:2019imo,Gralla:2019xty} for a discussion on this point. Since the hairy BHs in the region of interest are not very hairy, it is conceivable this offset is of a similar order.
\item[$\bullet$] The gas data~\cite{Walsh:2013uua} was included in our discussion for completeness, as it was in the EHT paper VI~\cite{Akiyama:2019eap}. However, this data is under tension even with the Kerr hypothesis, as discussed in detail in~\cite{Akiyama:2019eap}.  If the gas observations were to hold, they would have major implications concerning the Kerr paradigm. The conclusion that could be extracted here from this data is not different from the EHT paper: it is in tension with the models that were considered (including Kerr). 
\end{description}}

It would be very interesting to repeat the current analysis for the case of BHs with ultralight synchronised vector hair or scalar hair with self-interactions.

\section*{Acknowledgements}

P.C.  is  supported  by Grant No.  PD/BD/114071/2015 under the FCT-IDPASC Portugal Ph.D. program. This work is supported by the Funda\c{c}\~ao para a Ci\^encia e a Tecnologia (FCT) project UID/MAT/04106/2019 (CIDMA), by CENTRA (FCT) strategic project UID/FIS/00099/2013, by national funds (OE), through FCT, I.P., in the scope of the framework contract foreseen in the numbers 4, 5 and 6 of the article 23, of the Decree-Law 57/2016, of August 29,
changed by Law 57/2017, of July 19. We acknowledge support  from the project PTDC/FIS-OUT/28407/2017. This work has further been supported by  the  European  Union's  Horizon  2020  research  and  innovation  (RISE) programmes H2020-MSCA-RISE-2015 Grant No.~StronGrHEP-690904 and H2020-MSCA-RISE-2017 Grant No.~FunFiCO-777740. The authors would like to acknowledge networking support by the
COST Action CA16104.


\bibliography{Ref}{}  
\bibliographystyle{ieeetr}

\end{document}